\documentclass[preprint,showpacs,showkeys,preprintnumbers,aps,amssymb]{revtex4}

\usepackage{graphicx}
\usepackage{dcolumn}
\usepackage{bm}

\renewcommand{\theequation}%
     {\thesection.\arabic{equation}}

\def\ben{\begin{equation}}
\def\een{\end{equation}}
\def\bey{\begin{eqnarray}}
\def\eey{\end{eqnarray}}
\def\ba{\begin{array}}
\def\ea{\end{array}}
\def\benmrt{\begin{enumerate}}
\def\eenmrt{\end{enumerate}}

\def\psla{p{\raise1pt\hbox{$\!\!/$}}}
\def\dsla{\partial{\raise1pt\hbox{$\!\!\!/$}}}
\def\Dsla{D{\raise1pt\hbox{$\!\!\!/$}}}
\def\xsla{x{\raise1pt\hbox{$\!\!\!/$}}}

\def\jmu5{j_{\mu 5}^{(i)}(0)}
\def\jnu5{j_{\nu 5}^{(i)}(0)}

\def\qq0v{\langle0\!\mid\!{\bar q}q\!\mid\! 0\rangle}

\def\qc0f{\langle0\!\mid\!{\bar q}q\!\mid\!0\rangle_{F}}

\def\qsq0f{\langle0\!\mid\!{\bar q}\sigma_{\mu\nu}q\!\mid\!0\rangle_{F}}

\def\qgdq0f{\langle0\!\mid\!{\bar q}{\cal 
S}\gamma_{\mu}D_{\nu}q\!\mid\!0\rangle_{F}}

\def\qddq0f{\langle0\!\mid\!{\bar q}{\cal
S}D_{\mu}D_{\nu}q\!\mid\!0\rangle_{F}}

\def\3mmtm{|{\bf q}|^2}

\def\eq#1{Eq.(\ref{#1})}
\def\eqs#1#2{Eqs.(\ref{#1}) and (\ref{#2})}
\def\eqst#1#2#3{Eqs.(\ref{#1}), (\ref{#2}) and (\ref{#3})}

\def\Ref#1{[\ref{#1}]}
\def\Refs#1#2{[\ref{#1},\ref{#2}]}
\def\refref#1#2{[\ref{#1}-\ref{#2}]}

\def\p0{p_0}

\def\gam3{\mbox{\boldmath{$\gamma$}}}
\def\e0{E_{0}(s_{0},s)}
\def\e1{E_{1}(s_{0},s)}
\def\e2{E_{2}(s_{0},s)}

\def\aplt{\kern0.3333em \raise 0.2ex \hbox{$<$}%
\kern-0.8em \lower0.8ex \hbox{$\sim$}%
\kern0.3333em}

\def\absk{|{\bf k}|}
\def\absp{|{\bf p}|}
\def\absq{|{\bf q}|}
\def\mesr#1{\frac{{\rm d}^4 #1}{(2\pi)^4}}

\def\prop#1#2{\frac{i}{#1^2-#2^2+i\eta}}

\def\ln{{\rm ln}}

\begin{document}

\preprint{KEK-TH-870}

\title{On the thermal sunset diagram for scalar field theories}

\author{Tetsuo NISHIKAWA}
 \email{nishi@post.kek.jp} 
\author{Osamu MORIMATSU}
 \email{osamu.morimatsu@kek.jp}
 \author{Yoshimasa HIDAKA}
 \email{hidaka@post.kek.jp}
\affiliation{%
Institute of Particle and Nuclear Studies, 
High Energy Accelerator Research Organization, 1-1, Ooho, 
Tsukuba, Ibaraki, 305-0801, Japan
}
%

\date{\today}

\begin{abstract}
We study the so-called  \lq\lq sunset diagram'', which is one of two-loop self-energy diagrams, for scalar field theories at finite temperature. 

For this purpose, we first find the complete expression of the bubble diagram, the one-loop subdiagram of the sunset diagram, for arbitrary momentum. The temperature-dependent discontinuous part as well as the well-known temperature independent part is obtained analytically.
The continuous part  is reduced to a one-dimensional integral which one can easily evaluate numerically.

We calculate the temperature independent part and dependent part of the sunset diagram separately.
For the former, we obtain the discontinuous part first and the finite continuous part next using a twice-subtracted  dispersion relation. For the latter, we express it as a one-dimensional integral in terms of the bubble diagram. 

We also study the structure of the discontinuous part of the sunset diagram.
Physical processes, which are responsible for it, are identified. Processes due to the scattering with particles in the heat bath exist only at finite temperature and generate discontinuity for arbitrary momentum, which is a remarkable feature of the two-loop diagrams at finite temperature.
 
As an application of our result, we study the effect of the diagram on the spectral function 
of the sigma meson at finite temperature in the linear sigma model, which was obtained at one-loop order previously. At high temperature where the decay $\sigma\rightarrow\pi\pi$ is forbidden,
sigma acquires a finite width of the order of  $10\,{\rm MeV}$ while within the one-loop calculation its width vanishes. At low temperature, the spectrum does not deviate much from that at one-loop order. 
Possible consequences with including other two-loop diagrams are discussed.
\end{abstract}
\maketitle

\newpage
\section{Introduction}
\label{intro}
\setcounter{equation}{0}
The so-called sunset diagram, which is depicted in Fig.1, 
is one of two-loop self-energy diagrams.
The sunset diagram appears in theories with 4-point vertices 
such as ${\cal O}(4)$ linear sigma model, $\phi^{4}$ theory, and so on.

\begin{figure}[b]
\begin{center}
\includegraphics{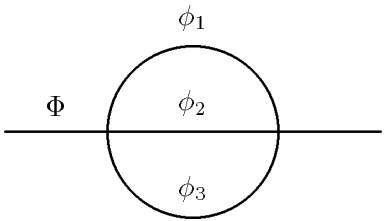}
\caption{\lq\lq Sunset'' diagram. We label the external particle by 
\lq\lq$\Phi$'' and internal particles with mass $m_i$ by $\phi_i$.}
\label{sunset}
\end{center}
\end{figure}

At finite temperature, two-loop self-energy diagrams have 
remarkable features that are not seen at zero temperature.
Consider the discontinuity of the self-energy.
At zero temperature, the discontinuity of one-loop diagrams is due to two particle intermediate states 
and that of two-loop diagrams comes from three particle states in addition to two particle ones.
In general, in higher loop diagrams new processes appear. 
However, they contribute to the discontinuity only at higher energies.
On the other hand, at finite temperature, even though the number of particles participating in the process at a given order of loops is the same as that at zero temperature, 
new processes appear even at low energy.
This is because at finite temperature some of particles participating in the process can be particles in the heat bath.
Furthermore, as will be shown in this paper, 
there exist processes which are possible at arbitrary energy.
Accordingly, the discontinuity of two-loop self-energies is non-vanishing in all the energy region.
Therefore, in some cases at finite temperature, 
to extend calculations to two-loop order has a meaning more than just making more precision.

One of such cases is the spectral function for the sigma meson in the linear sigma model.
The spectral function of the sigma meson at finite temperature was studied
by Chiku and Hatsuda at one-loop level \Ref{chiku}.
The sigma meson at zero temperature has a large width due to the strong coupling 
with two pions.
However, they found that at finite temperature the spectral function 
near $\sigma\rightarrow\pi\pi$ threshold is enhanced
as a typical signal of chiral phase transition.
This is because, as the temperature increases, the mass of $\sigma$ decreases 
while that of $\pi$ increases due to partial restoration of chiral symmetry
and, accordingly, the phase space available for 
the $\sigma\rightarrow\pi\pi$ decay is squeezed to zero.
At finite temperature, however, there exist processes, collision with or absorption of a thermal particle in the heat bath,
which contributes to the discontinuity at arbitrary energy for $\sigma$, as discussed above.
If we include them,
the structure of the spectral function might be significantly modified.
Their effects on the spectral function 
cannot be taken into account
till we extend the calculation including two-loop self-energies.

In this paper, we take a step forward in the calculation of two-loop self-energy diagrams, i.e. we evaluate the thermal sunset diagram for scalar field theories.
After giving a brief review of the real time formalism 
\refref{schwin}{LeBellac} in the second section,
which we use for the calculation of diagrams, 
we first examine the bubble diagram, a one-loop diagram which appears as a subdiagram of the sunset diagram, at finite temperature in the third section.
Then, we discuss the structure of the discontinuous part
of the thermal sunset diagram in the fourth section
and explain how we calculate the sunset diagram in the fifth section. 
Using our result for this diagram,
we study the effects of the diagram on the spectral function 
of $\sigma$ at finite temperature 
in ${\cal O}(4)$ linear sigma model in the sixth section. 
Finally we summarize the paper in the seventh section.

\section{Brief review of the real time formalism}
\label{rtf}
For the calculations of thermal Feynman diagrams we adopt the real time formalism
throughout this paper.
We briefly review the formalism in this section.

In the real time formalism propagators are given by $2\times2$ matrices.
The 4-components of the free propagator of a scalar particle with mass $m$
are given by [\ref{LeBellac}]
\bey
i\Delta_{11}^{F}(k;m)&=&
\frac{i}{k^2-m^2+i\eta}+n(k_0)2\pi\delta(k^2-m^2),\\
i\Delta_{12}^{F}(k;m)&=&
e^{\sigma k_0}[n(k_0)+\theta(-k_0)]2\pi\delta(k^2-m^2),\\
i\Delta_{21}^{F}(k;m)&=&
e^{-\sigma k_0}[n(k_0)+\theta(k_0)]2\pi\delta(k^2-m^2),\\
i\Delta_{22}^{F}(k;m)&=&
\frac{-i}{k^2-m^2-i\eta}+n(k_0)2\pi\delta(k^2-m^2),
\label{freepro}
\eey
where $n(k_0)$ is the Bose-Einstein distribution function at temperature 
$T\equiv 1/\beta$:
\ben
n(k_0)=\frac{1}{{\rm exp}(\beta|k_0|)-1}.
\een 
Off-diagonal elements depend on a free parameter $\sigma$.
In this paper we make the symmetrical choice $\sigma=\beta/2$,
leading to
\bey
i\Delta_{12}^{F}(k;m)&=&i\Delta_{21}^{F}(k;m)
=e^{\beta|k_0|/2}n(k_0)2\pi\delta(k^2-m^2).
\eey

We denote the matrix of the full propagator by
\bey
\tilde{\Delta}(k;m)=
\left(
\begin{array}{cc}
\Delta_{11}(k;m)&\Delta_{12}(k;m)\\
\Delta_{21}(k;m)&\Delta_{22}(k;m)
\end{array}
\right).
\eey
Each component is the solution of the Schwinger-Dyson equation:
\bey
\Delta_{ab}(k;m)=\Delta_{ab}^{F}(k;m)+\Delta_{ac}^{F}(k;m)\Pi_{cd}(k)\Delta_{db}(k;m),
\label{SD}
\eey
where $\Pi_{cd}(k)$ is the self-energy.

The following matrix
\ben
U(k)=\left(
\begin{array}{cc}
\sqrt{1+n(k_0)}&\sqrt{n(k_0)}\\
\sqrt{n(k_0)}&\sqrt{1+n(k_0)}
\end{array}
\right),
\een
\lq diagonalizes' the free propagator:
\ben
U(k)^{-1}\left(
\begin{array}{cc}
\Delta_{11}^{F}(k;m)&\Delta_{12}^{F}(k;m)\\
\Delta_{21}^{F}(k;m)&\Delta_{22}^{F}(k;m)
\end{array}
\right)U(k)^{-1}=
\left(
\begin{array}{cc}
\Delta_{0}^{F}(k;m)&0\\
0&-\Delta_{0}^{F}(k;m)^*
\end{array}
\right),
\een
where $\Delta_{0}^{F}(k;m)$ is the Feynman propagator at $T=0$
\ben
\Delta_{0}^{F}(k;m)=\frac{1}{k^2-m^2+i\eta}.
\een
It is known that $U(k)$ also diagonalizes the full propagator
[\ref{LeBellac}] as well as $\Pi_{ab}(k)$
\ben
\Pi_{ab}(k)=(U(k)^{-1})_{ac}\left(
\begin{array}{cc}
\bar{\Pi}(k)&0\\
0&-\bar{\Pi}(k)^*
\end{array}
\right)_{cd}(U(k)^{-1})_{db}.
\een
This matrix equation gives relations between matrix elements:
\bey
{\rm Re}\bar{\Pi}(k)&=&{\rm Re}\Pi_{11}(k),\label{rel-SE1}\\
{\rm Im}\bar{\Pi}(k)&=&\tanh(\beta|k_0|/2){\rm Im}\Pi_{11}(k),\label{rel-SE2}\\
{\rm Im}\bar{\Pi}(k)&=&i\sinh(\beta|k_0|/2)\Pi_{12}(k).\label{rel-SE3}
\eey

Let us next find the expression of the spectral function.
$\Delta_{11}$ has the following spectral representation:
\ben
\Delta_{11}(k;m)=\int{\rm d}\omega^2 \rho(\omega,{\bf k})
\Delta_{11}^F(k_0,\omega),
\label{spe-delta11}
\een
where we denote the argument of $\Delta_{11}^F$ 
by $k_0$ and $\omega\equiv\sqrt{{\bf k}^2+m^2}$
instead of $k$ and $m$.
One can prove that, if $\Delta_{11}$ is given in \eq{spe-delta11},
the Green function 
\ben
\Delta(t,{\bf x})=\frac{i}{(2\pi)^4}
\int{\rm d}^4 k e^{ik\cdot x}\Delta_{11}(k;m)
\een
obeys the Kubo-Martin-Schwinger (KMS) condition [\ref{sem-ume}].
From \eq{spe-delta11} we obtain
\ben
\rho(k)=-\frac{1}{\pi}\tanh(\beta|k_0|/2){\rm Im}\Delta_{11}(k;m).
\label{rho-Imdelta11}
\een
Our next task is thus to find the expression of ${\rm Im}\Delta_{11}(k;m)$.
Denoting the diagonalized full propagator by
\bey
U(k)^{-1}\tilde{\Delta}(k;m)U(k)^{-1}&=&
\left(
\begin{array}{cc}
\Delta(k;m)&0\\
0&-\Delta(k;m)^{*}
\end{array}
\right),
\eey
we obtain
\bey
\tilde{\Delta}&=&
U(k)\left(
\begin{array}{cc}
\Delta(k;m)&0\\
0&-\Delta(k;m)^{*}
\end{array}
\right)
U(k)\cr
&=&\left(
\begin{array}{cc}
[1+n(k_0)]\Delta(k;m)-n(k_0)\Delta(k;m)^{*}&
\sqrt{n(k_0)[1+n(k_0)]}[\Delta(k;m)-\Delta(k;m)^{*}]\\
\sqrt{n(k_0)[1+n(k_0)]}[\Delta(k;m)-\Delta(k;m)^{*}]
&-[1+n(k_0)]\Delta(k;m)^{*}+n(k_0)\Delta(k;m)
\end{array}
\right).\cr
&&
\eey
Taking the $(1,1)$-components in the both sides of the above equation
yields
\ben
{\rm Im}\Delta_{11}(k;m)=\coth(\beta|k_0|/2){\rm Im}\Delta(k;m).
\label{imdelta11-imdelta}
\een
The expression for $\Delta(k;m)$ can be found in the following manner.
In \eq{SD}, iteratively using \eq{SD} itself in the right hand side
and diagonalizing, we obtain
\bey
&&U(k)^{-1}\tilde{\Delta}(k;m)U(k)^{-1}=
\left(
\begin{array}{cc}
\Delta_{0}^{F}(k;m)&0\\
0&-\Delta_{0}^{F}(k;m)^{*}
\end{array}
\right)\cr
&&+
\left(
\begin{array}{cc}
\Delta_{0}^{F}(k;m)&0\\
0&-\Delta_{0}^{F}(k;m)^{*}
\end{array}
\right)
\left(
\begin{array}{cc}
\bar{\Pi}(k)&0\\
0&-\bar{\Pi}(k)^*
\end{array}
\right)
\left(
\begin{array}{cc}
\Delta_{0}^{F}(k;m)&0\\
0&-\Delta_{0}^{F}(k;m)^{*}
\end{array}
\right)+\cdots.\cr
&&
\eey
The (1,1) component of this equation gives
\bey
\Delta(k;m)&=&(U(k)^{-1}\tilde{\Delta}(k;m)U(k)^{-1})_{11}\cr
&=&\Delta_{0}^{F}(k;m)+\Delta_{0}^{F}(k;m)\bar{\Pi}(k)\Delta_{0}^{F}(k;m)
+\cdots\cr
&=&\frac{1}{k^2-m^2-\bar{\Pi}(k)}.
\label{delta}
\eey
Therefore, from \eqst{rho-Imdelta11}{imdelta11-imdelta}{delta},
we obtain the desired expression 
of the spectral function:
\ben
\rho(k)=-\frac{1}{\pi}
\frac{{\rm Im}\bar{\Pi}(k)}{[k^2-m^2-{\rm Re}\bar{\Pi}(k)]^2
+[{\rm Im}\bar{\Pi}(k)]^2}.
\label{rho-RTF}
\een
Thus the spectral function can be written only with one component
of the self-energies
while all the components enter $\Delta_{11}(k;m)$
(see \eq{SD}).

\section{Bubble diagram}
In this section we examine the \lq\lq bubble" diagram, which is a one-loop diagram shown in Fig.2.
The bubble diagram appears as a subdiagram of the sunset diagram and therefore we need the expression of the former in the calculation of the latter.

\begin{figure}[h]
\begin{center}
\includegraphics{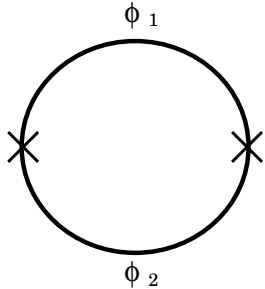}
\caption{\lq\lq Bubble'' diagram.}
\label{bubble}
\end{center}
\end{figure}

Among the 4-components of the bubble diagram we need only the (1,1) component
since the sunset diagram does not have any internal vertices.
The (1,1) component of the bubble diagram for scalar particles is given by
\ben
{\cal I}_{\rm bub}(p;m_1,m_2)_{11}=\int\mesr{k}
i\Delta_{11}^{F}(p+k;m_1)i\Delta_{11}^{F}(k;m_2).\label{1loop}
\een
\eq{1loop} can be expressed as the sum of terms with different numbers of 
Bose-Einstein factors, $n$:
\ben
{\cal I}_{\rm bub}(p;m_1,m_2)_{11}=I^{(2)}(p^2;m_1,m_2)
+(F^{(2)}(p;m_1,m_2)+(1\leftrightarrow 2))+F^{(3)}(p;m_1,m_2).
\een
Each term in the right hand side is respectively given by
\bey
I^{(2)}(p^2;m_1,m_2)&=&\int\mesr{k}
\frac{i}{(p+k)^2-m_1^2+i\eta}\frac{i}{k^2-m_2^2+i\eta},
\label{I2}\\
F^{(2)}(p;m_1,m_2)&=&\int\mesr{k}\frac{i}{(p+k)^2-m_1^2+i\eta}n(k_0)2\pi\delta(k^2-m_2^2),
\label{F2}
\\
F^{(3)}(p;m_1,m_2)&=&\int\mesr{k}n(p_0+k_0)2\pi\delta((p+k)^2-m_1^2)n(k_0)2\pi\delta(k^2-m_2^2).
\label{F3}
\eey
Let us carry out the integration in the above equations.
The $T$-independent part $I^{(2)}$ is given in textbooks [\ref{ramond}]. 
In $d$-dimension it is given by
\bey
I^{(2)}(p^2;m_1,m_2)&=&\frac{i}{16\pi^2}
\left(-\frac{1}{\bar\epsilon}+x_{+}\ln\frac{m_1^2}{\kappa^2}
-x_{-}\ln\frac{m_2^2}{\kappa^2}-2-I\right),
\label{I2-result}
\eey
where $\kappa$ is the renormalization point
and $\frac{1}{\bar\epsilon}$ and $x_\pm$ are 
\ben
\frac{1}{\bar\epsilon}\equiv\frac{1}{\epsilon}
-\gamma+{\rm ln}4\pi\quad(\epsilon=(4-d)/2,
\quad\gamma: {\rm Euler\,\,constant}),
\een
\bey
x_\pm\equiv\pm\frac{1}{2}+\frac{m_1^2-m_2^2}{2p^2}.
\eey
$I$ is given by
\bey
I&=&\left\{
\begin{array}{ll}
\sqrt{C}\left[\ln\frac{(x_{+}-\sqrt{C})(x_{-}+\sqrt{C})}
                       {(x_{-}-\sqrt{C})(x_{+}+\sqrt{C})}
               +i 2\pi\right]
&{\rm for}\quad p^2>(m_1+m_2)^2,\\
-2\sqrt{D}\left[
{\rm arctan}\frac{x_+}{\sqrt{D}}-{\rm arctan}\frac{x_-}{\sqrt{D}}\right]&
{\rm for}\quad(m_1-m_2)^2<p^2<(m_1+m_2)^2,\\
\sqrt{C}\ln\frac{(\sqrt{C}-x_{+})(\sqrt{C}+x_{-})}
                {(\sqrt{C}-x_{-})(\sqrt{C}+x_{+})}
&{\rm for}\quad p^2<(m_1-m_2)^2,
\end{array}
\right.\cr
&&
\label{I}
\eey
where $C$ and $D$ are
\bey
C&\equiv&\frac{1}{4}
\left[1-\frac{(m_1+m_2)^2}{p^2}\right]\left[1-\frac{(m_1-m_2)^2}{p^2}\right],\quad
D\equiv-C.
\eey
We see from \eqs{I2}{I} that the discontinuous part, ${\rm Im}iI^{(2)}$,
is non-vanishing for $p^2>(m_1+m_2)^2$.

The divergent part in \eq{I2} is renormalized by applying a counter term 
to the Lagrangian. The finite part is determined in such a way 
that the resultant self-energy satisfies a proper normalization condition. 

Let us next turn to the $T$-dependent part, 
$F^{(2)}$ and $F^{(3)}$.
We first discuss their discontinuous parts.
As shown in Appendix \ref{cond-disc} the discontinuous part of $F^{(2)}(p;m_1,m_2)$ is given analytically as follows:
\benmrt
\item
For $p^2>(m_1+m_2)^2$ or $0<p^2<(m_1-m_2)^2$,
\bey
{\rm Im}iF^{(2)}(p;m_1,m_2)
&=&\frac{1}{16\pi|{\bf p}|}
\frac{1}{\beta}
\ln\left|\frac{1-{\rm e}^{-\beta\omega_+}}
{1-{\rm e}^{-\beta\omega_-}}\right|.
\label{ImiF2-1}
\eey
\item
For $(m_1-m_2)^2<p^2<(m_1+m_2)^2$,
\bey
{\rm Im}iF^{(2)}(p;m_1,m_2)&=&0.
\label{ImiF2-2}
\eey
\item
For $p^2<0$,
\bey
{\rm Im}iF^{(2)}(p;m_1,m_2)
&=&\frac{1}{16\pi|{\bf p}|}\frac{-1}{\beta}
\ln\left|(1-{\rm e}^{-\beta\omega_+})(1-{\rm e}^{-\beta\omega_-})\right|.
\label{ImiF2-3}
\eey
\eenmrt
Here, 
\ben
\omega_{\pm}
=\frac{1}{2}\left|
\sqrt{\left(1+\frac{m_2^2-m_1^2}{p^2}\right)^2}|p_0|
\pm\sqrt{\left[1-\frac{(m_2+m_1)^2}{p^2}\right]
\left[1-\frac{(m_1-m_2)^2}{p^2}\right]}\absp\right|.
\label{omega}
\een

$F^{(3)}$ has only discontinuous part.
Its calculation can be done 
in a way similar to that for
${\rm Im}iF^{(2)}(p;m_1,m_2)$.
We show the final results:
\benmrt
\item
For $p^2>(m_1+m_2)^2$,
\bey
F^{(3)}(p;m_1,m_2)
&=&\frac{1}{8\pi\absp\beta}\frac{1}{{\rm e}^{\beta|p_0|}-1}
{\rm ln}\left|\frac{1-{\rm e}^{-\beta\omega_+}}{1-{\rm e}^{-\beta\omega_-}}
\frac{{\rm e}^{\beta(|p_0|-\omega_-)}-1}
     {{\rm e}^{\beta(|p_0|-\omega_+)}-1}\right|.
\label{F3-1}
\eey
\item
For $(m_1-m_2)^2<p^2<(m_1+m_2)^2$,
\bey
F^{(3)}(p;m_1,m_2)&=&0.
\label{F3-2}
\eey
\item
For $0<p^2<(m_1-m_2)^2$,
\bey
F^{(3)}(p;m_1,m_2)
&=&\frac{1}{8\pi\absp\beta}\cdot\frac{1}{{\rm e}^{\beta|p_0|}-1}\cr
&&\times\left[
{\rm ln}\left|\frac{1-{\rm e}^{-\beta\omega_+}}
                   {1-{\rm e}^{-\beta\omega_-}}\right|
-{\rm e}^{\beta|p_0|}
{\rm ln}\left|\frac{1-{\rm e}^{-\beta(|p_0|+\omega_+)}}{1-{\rm e}^{-\beta(|p_0|+\omega_-)}}\right|
\right].
\label{F3-3}
\eey
\item
For $p^2<0$,
\bey
F^{(3)}(p;m_1,m_2)
&=&\frac{1}{8\pi\absp\beta}\left\{\frac{1}{e^{-\beta|p_0|}-1}
\left[
-{\rm ln}|1-{\rm e}^{-\beta\omega_+}|
+{\rm e}^{-\beta|p_0|}{\rm ln}|1-{\rm e}^{-\beta(\omega_{+}-|p_0|)}|
\right]\right.\cr
&&
\left.
+\frac{1}{e^{\beta|p_0|}-1}\left[
-{\rm ln}|1-{\rm e}^{-\beta\omega_-}|
+{\rm e}^{\beta|p_0|}{\rm ln}|1-{\rm e}^{-\beta(\omega_{-}+|p_0|)}|
\right]
\right\}.\cr
&&
\label{F3-4}
\eey
\eenmrt
In the above equations $\omega_{\pm}$ are given by \eq{omega} 
with $m_1$ and $m_2$ replaced by ${\rm max}\{m_1,m_2\}$ 
and ${\rm min}\{m_1,m_2\}$, respectively,
since $F^{(3)}(p;m_1,m_2)$ is symmetric with respect to $m_1$ and $m_2$ 
by definition.

It should be noted that the $T$-dependent part has two cuts 
in the complex $p^2$ plane: one starts from $p^2=(m_1+m_2)^2$
to the right along the real axis and the other
from $p^2=(m_1-m_2)^2$ to the left.

Let us next discuss the continuous part of $F^{(2)}$,
which is
\bey
{\rm Re}iF^{(2)}(p;m_1,m_2)
=-\int\mesr{k}{\cal P}\frac{1}{(p+k)^2-m_1^2}n(k_0)2\pi\delta(k^2-m_2^2),
\label{reiF2}
\eey
where ${\cal P}$ stands for the prescription of Cauchy's principal value.
After integration over $k_0$ and angle, we can express \eq{reiF2} as a one-dimensional integral:
\bey
&&{\rm Re}iF^{(2)}(p;m_1,m_2)\cr
&=&\frac{1}{16\pi^2 |{\bf p}|}\int_{m_2}^{\infty}{\rm d}\omega n(\omega)
\cr
&\times&
{\cal P}\ln\left|
\frac{(p^2+2p_0 \omega-2|{\bf p}|\sqrt{\omega^2-m_2^2}+m_2^2-m_1^2)
      (p^2-2p_0 \omega-2|{\bf p}|\sqrt{\omega^2-m_2^2}+m_2^2-m_1^2)}
     {(p^2+2p_0 \omega+2|{\bf p}|\sqrt{\omega^2-m_2^2}+m_2^2-m_1^2)
      (p^2-2p_0 \omega+2|{\bf p}|\sqrt{\omega^2-m_2^2}+m_2^2-m_1^2)}\right|,
\cr
&&
\label{cont-F2}
\eey
whose integration will be carried out numerically.

\section{Structure of the discontinuous part of the sunset diagram}
\label{structure-impart}
Before proceeding to the calculation of the thermal sunset diagram,
we study the structure of the discontinuous part of 
the diagram.
In ref.[\ref{weldon}], Weldon analyzed the discontinuous part of the bubble diagram.
We will generalize it for the sunset diagram.
For this purpose it is convenient
to use \eq{rel-SE3} [\ref{fujimoto}], namely
\bey
&&{\rm Im}i\bar{\cal I}_{\rm sun}(k;m_1,m_2,m_3)
=i\sinh(\beta|k_0|/2)i{\cal I}_{\rm sun}(k;m_1,m_2,m_3)_{12}.
\label{sunsetim}
\eey
Here $\bar{\cal I}_{\rm sun}(k;m_1,m_2,m_3)$ is the $(1,1)$ component of 
the diagonalized self-energy matrix for the sunset diagram.
We note that the self-energy which actually enters spectral functions 
is diagonalized one (see \eq{rho-RTF}).
${\cal I}_{\rm sun}(k;m_1,m_2,m_3)_{12}$ is the $(1,2)$
component of the sunset diagram,
which is given by the following integral:
\bey
i{\cal I}_{\rm sun}(k;m_1,m_2,m_3)_{12}&=&\int\mesr{p}i\Delta_{12}^F(p;m_1)
i{\cal I}_{\rm bub}(k-p;m_2,m_3)_{12}.
\label{sunset12}
\eey
For $i\Delta_{12}^F(p;m_1)$ it is convenient to use another form:
\bey
i\Delta_{12}^F(p;m_1)&=&e^{\beta p_0/2}f(p_0)
\epsilon(p_0)2\pi\delta(p^2-m_1^2),\cr
f(p_0)&=&\frac{1}{e^{\beta p_0}-1}.
\eey
$i{\cal I}_{\rm bub}(p;m_1,m_2)_{12}$ denotes 
the $(1,2)$ component of the bubble diagram:
\bey
i{\cal I}_{\rm bub}(k-p;m_2,m_3)_{12}
&=&i\int\mesr{q}i\Delta_{12}^F(q;m_2)i\Delta_{12}^F(k-p-q;m_3).
\eey
One can reduce this equation to
\bey
&&i{\cal I}_{\rm bub}(p;m_1,m_2)_{12}\cr
&=&i e^{\beta(k_0-p_0)/2}f(k_0-p_0)\cr
&&\times\int\frac{{\rm d}^4 q}{(2\pi)^2}
\frac{1}{4E_2 E_3}
\left\{(1+n_2+n_3)
\left[\delta(k_0-p_0-E_2-E_3)-\delta(k_0-p_0-E_2+E_3)\right]
\right.\cr
&&\left.
-(n_2-n_3)
\left[\delta(k_0-p_0-E_2+E_3)-\delta(k_0-p_0+E_2-E_3)\right]
\right\},
\label{1loop12}
\eey
where $E_2=\sqrt{\absq^2+m_2^2}$ and $E_3=\sqrt{|{\bf k-p-q}|^2+m_3^2}$. 
We also define $E_1=\sqrt{\absp^2+m_1^2}$ for later use.
$n_i$ is the Bose-Einstein factor defined by $n_i=n(E_i)$.
Using \eq{sunsetim} and \eq{sunset12} in which 
\eq{1loop12} is substituted yields
\bey
&&{\rm Im}i\bar{\cal I}_{\rm sun}(k;m_1,m_2,m_3)\cr
&=&-\pi\epsilon(k_0)\int\frac{{\rm d}^3 p}{(2\pi)^3}
\int\frac{{\rm d}^3 q}{(2\pi)^3}\frac{1}{8E_1 E_2 E_3}\cr
&&\times\left\{
((1+n_1)(1+n_2)(1+n_3)-n_1 n_2 n_3)\delta(k_0-E_1-E_2-E_3)
\right.\cr
&&\quad
+(n_1 n_2 n_3-(1+n_1)(1+n_2)(1+n_3))\delta(k_0+E_1+E_2+E_3)\cr
&&\quad
+[(n_2 n_3 (1+n_1)-n_1 (1+n_2)(1+n_3))\delta(k_0-E_1+E_2+E_3)\cr
&&
\left.
\quad
+(n_1 (1+n_2)(1+n_3)-n_2 n_3 (1+n_1))\delta(k_0+E_1-E_2-E_3)
+({\rm permutations})]
\right\}.\cr
&&
\label{imsunset2}
\eey
In this equation \lq\lq permutations'' stands for the terms obtained 
by permuting the particle labels of the third and the forth terms.

\begin{figure}[!b]
\begin{center}
\includegraphics{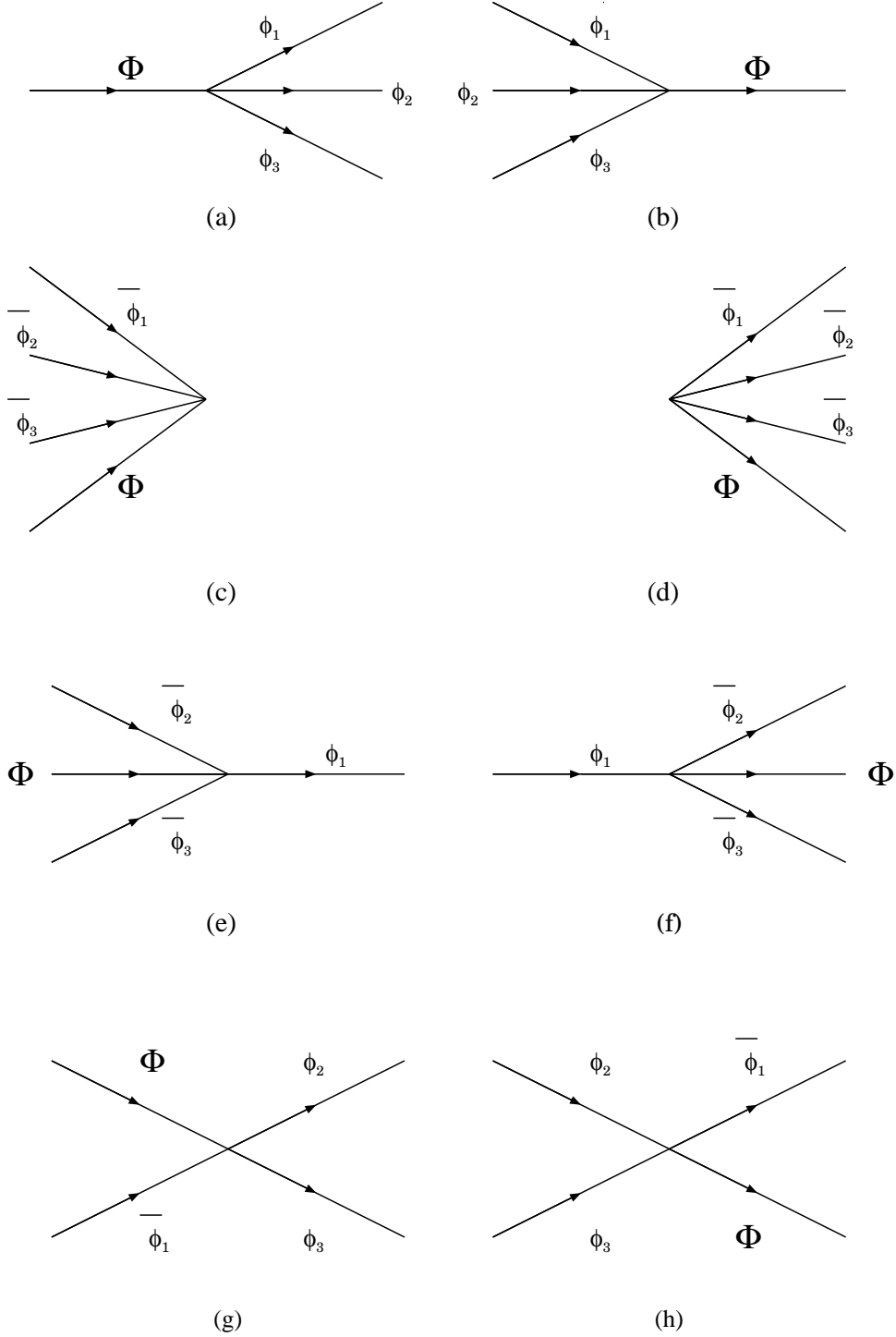}
\caption{The amplitudes in \eq{imsunset2} responsible for the disappearance
and reappearance of $\Phi$. 
${\bar\phi}_i$ stands for an anti-particle of $\phi_i$.
(a) minus (b) corresponds to the first term in \eq{imsunset2}, 
(c) minus (d) to the second, (e) minus (f) to the third 
and (g) minus (h) to the forth. Amplitudes obtained by permuting
particle labels of (e), (f), (g) and (h) also exist.}
\label{decaysunset}
\end{center}
\end{figure}

Let us now consider the physical content of \eq{imsunset2}.
The first term in \eq{imsunset2} may be interpreted as the probability
for the decay $\Phi\rightarrow \phi_1\phi_2\phi_3$
with the statistical weight $(1+n_1)(1+n_2)(1+n_3)$
for stimulated emission 
minus the probability for the creation $\phi_1\phi_2\phi_3\rightarrow \Phi$
with the weight $n_1 n_2 n_3$ for absorption.
The second term is the anti-particle counter part 
of the first term.
The third term represents the probability for
$\Phi{\bar\phi}_2{\bar\phi}_3\rightarrow\phi_1$
with the weight $n_2 n_3 (1+n_1)$ minus that for 
$\phi_1\rightarrow\Phi{\bar\phi}_2{\bar\phi}_3$ with the weight 
$n_1 (1+n_2)(1+n_3)$. 
Here ${\bar\phi}_i$ stands for the anti-particle of $\phi_i$.
The forth term is the anti-particle counter part
of the third term.
It represents the probability for $\Phi{\bar\phi}_1\rightarrow\phi_2\phi_3$
with the weight $n_1 (1+n_2) (1+n_3)$ minus that for
$\phi_2\phi_3\rightarrow\Phi{\bar\phi}_1$ with the weight 
$n_2 n_3 (1+n_1)$.
All processes are shown in Fig.\ref{decaysunset}.

We next find the region of $k^2$ 
where the physical processes contained in \eq{imsunset2} are possible,
which is equivalent to looking for the condition under which the integral 
over q in \eq{imsunset2} survives.
For the first and the second terms in \eq{imsunset2} 
to be non-vanishing, $k^2$ must satisfies the condition $k^2>(m_1+M_{23})^2$, 
where $M_{23}$ is the invariant mass 
of $\phi_2$ and $\phi_3$.
Therefore, the processes in Fig.\ref{decaysunset} (a), (b), (c) and (d)
are possible for $k^2>(m_1+m_2+m_3)^2$ since $M_{23}>m_2+m_3$. 
The third and the forth terms survive when $k^2<(m_1-M_{23})^2$.
The processes in Fig.\ref{decaysunset} (e), (f), (g) and (h),
therefore, take place at arbitrary $k^2$.
This is reasonable since the processes 
in Fig.\ref{decaysunset} (g) and (h) are scattering ones 
and since those in Fig.\ref{decaysunset} (e) and (f) can be also regarded 
as scattering 
by interpreting the incoming $\Phi$ in (e) as an outgoing anti-$\Phi$ 
and by doing the outgoing $\Phi$ in (f) as an incoming anti-$\Phi$.
Accordingly, the discontinuous part of the sunset diagram is non-vanishing for
arbitrary $k^2$, which is a remarkable feature of the thermal self-energy at and beyond two-loop order.

\section{Calculation of the thermal sunset diagram}

In this section, we explain how to calculate the thermal sunset diagram 
by reducing it to an expression written in terms of the bubble diagram 
previously obtained. 

The (1,1) component of the sunset diagram shown in Fig.\ref{sunset} is given by
\ben
{\cal I}_{\rm sun}(k;m_1,m_2,m_3)_{11}=
\int\mesr{p}
i\Delta_{11}^F(p;m_1)
\int\mesr{q}i\Delta_{11}^F(q;m_2)i\Delta_{11}^F(k-p-q;m_3).
\label{sunset0}
\een
We decompose \eq{sunset0} into terms without and with Bose-Einstein factors, which we denote by 
${\cal I}^{\rm vac}_{\rm sun}(k^2;m_1,m_2,m_3)_{11}$
and ${\cal I}^{\rm th}_{\rm sun}(k;m_1,m_2,m_3)_{11}$, respectively:
\ben
{\cal I}_{\rm sun}(k;m_1,m_2,m_3)_{11}
={\cal I}^{\rm vac}_{\rm sun}(k^2;m_1,m_2,m_3)_{11}
+{\cal I}^{\rm th}_{\rm sun}(k;m_1,m_2,m_3)_{11}.
\label{sunset0-1}
\een
They are given by
\bey
{\cal I}^{\rm vac}_{\rm sun}(k^2;m_1,m_2,m_3)_{11}
&=&\int\mesr{p}\prop{p}{m_1}I^{(2)}((k-p)^2;m_2,m_3),
\label{sunsetvac0}
\\
{\cal I}^{\rm th}_{\rm sun}(k;m_1,m_2,m_3)_{11}
&=&\int\mesr{p}\prop{p}{m_1}\cr
&&\times
\left[\left(F^{(2)}(k-p;m_2,m_3)+(2\leftrightarrow 3)\right)
+F^{(3)}(k-p;m_2,m_3)\right]\cr
&+&\int\mesr{p}
n(p_0)2\pi\delta(p^2-m_1^2)\left[I^{(2)}((k-p)^2;m_2,m_3)\right.\cr
&&\left.+\left(F^{(2)}(k-p;m_2,m_3)+(2\leftrightarrow 3)\right)
+F^{(3)}(k-p;m_2,m_3)\right].\cr
&&
\label{sunset1}
\eey
Here we have expressed the second integral in \eq{sunset0} 
in terms of $I^{(2)}$, $F^{(2)}$ and $F^{(3)}$.
The $T$-independent part, \eq{sunsetvac0}, has a subdivergence coming from
the nested bubble diagram and a two-loop overall divergence.
On the other hand, $T$-dependent part, \eq{sunset1}, has only a subdivergence,
which can be removed by carrying out renormalization at one-loop level.
Hereafter, $I^{(2)}$ expresses that with the divergence removed.
We calculate \eqs{sunsetvac0}{sunset1} separately.

\subsection{$T$-independent part}

The $T$-independent part of sunset type diagrams
has been calculated by several authors \refref{post}{groote} so far.
In this paper we calculate its finite part using a dispersion relation.

We first find the discontinuous part of 
${\cal I}^{\rm vac}_{\rm sun}(k^2;m_1,m_2,m_3)_{11}$ and then
compute the finite continuous part using the obtained discontinuous part, {\it via} the twice-subtracted dispersion relation.

In order to calculate the discontinuous part of 
${\cal I}^{\rm vac}_{\rm sun}(k^2;m_1,m_2,m_3)_{11}$,
we substitute a dispersion relation 
for $iI^{(2)}((k-p)^2;m_2,m_3)$:
\bey
iI^{(2)}((k-p)^2;m_2,m_3)&=&\frac{1}{\pi}
\int_{(m_2+m_3)^2}^{\infty}{\rm d}M^2\frac{{\rm Im}iI^{(2)}(M^2;m_2,m_3)}
{M^2-(k-p)^2-i\eta},
\eey
into \eq{sunsetvac0}.
Then, we obtain
\bey
{\cal I}^{\rm vac}_{\rm sun}(k^2;m_1,m_2,m_3)_{11}
&=&\frac{1}{\pi}\int_{(m_2+m_3)^2}^{\infty}{\rm d}M^2{\rm Im}iI^{(2)}(M^2;m_2,m_3)\cr
&&\times\int\mesr{p}\prop{p}{m_1}\prop{(k-p)}{M}\cr
&=&\frac{1}{\pi}\int_{(m_2+m_3)^2}^{\infty}{\rm d}M^2{\rm Im}iI^{(2)}(M^2;m_2,m_3)
I^{(2)}(k^2;m_1,M).
\eey
We take the discontinuous part of this equation:
\bey
{\rm Im}i{\cal I}^{\rm vac}_{\rm sun}(k^2;m_1,m_2,m_3)_{11}
&=&\frac{1}{\pi}\int_{(m_2+m_3)^2}^{\infty}{\rm d}M^2{\rm Im}iI^{(2)}(M^2;m_2,m_3)
{\rm Im}iI^{(2)}(k^2;m_1,M).\cr
&&
\label{sunsetvac-im}
\eey
Using the expression for ${\rm Im}iI^{(2)}(p^2;m_,m_2)$ 
we can easily evaluate \eq{sunsetvac-im} numerically.

\begin{figure}[b]
\begin{center}
\includegraphics[width=15cm]{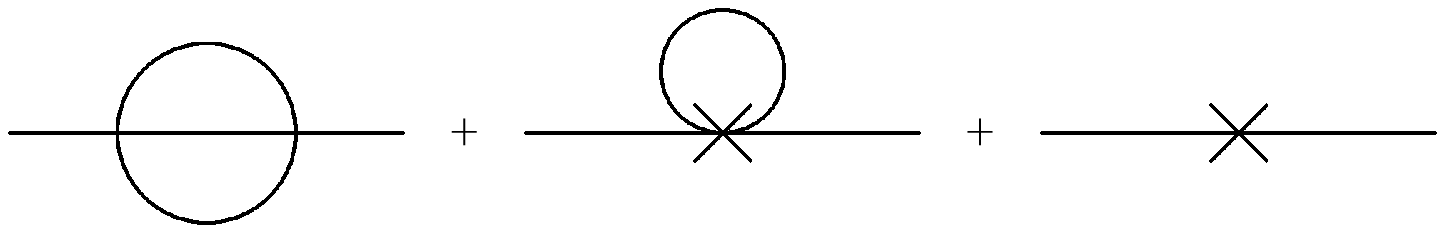}
\caption{Renormalized $T$-independent part of the sunset diagram.
The second and the third diagrams cancel the sub- and overall 
divergences in the first diagram (bare sunset diagram) 
respectively.}
\label{nthsunset}
\end{center}
\end{figure}

Next we turn to the continuous part.
The continuous part of \eq{sunsetvac0} is divergent.
Therefore one needs corresponding counter terms in the Lagrangian.
The second and third diagrams of Fig.\ref{nthsunset} appear
at two-loop order, which cancel the divergences in the bare sunset diagram, the first diagram of Fig.\ref{nthsunset}.
In this paper, we do not explicitly go through the renormalization procedure but concentrate on the finite part.

In order to calculate the finite part, we use the twice-subtracted dispersion relation:
\bey
&&{\rm Re}i\tilde{\cal I}^{\rm vac}_{\rm sun}(k^2;m_1,m_2,m_3)_{11}\cr
&\equiv&{\rm Re}\left\{i{\cal I}^{\rm vac}_{\rm sun}(k^2;m_1,m_2,m_3)_{11}
-i{\cal I}^{\rm vac}_{\rm sun}(0;m_1,m_2,m_3)_{11}-k^2 \left[\frac{\partial}{\partial k^2}i{\cal I}^{\rm vac}_{\rm sun}
(k^2;m_1,m_2,m_3)_{11}\right]_{k^2=0}
\right\}\cr
&=&\frac{k^4}{\pi}
\int^{\infty}_{(m_1+m_2+m_3)^2}{\rm d}M^2{\cal P}
\frac{{\rm Im}i{\cal I}^{vac}_{sunset}(M^2;m_1,m_2,m_3)}{M^4(M^2-k^2)}.
\label{sunsetvac-Re}
\eey
The second and the third terms in the left hand side are divergent and have to be renormalized
but the right hand side is finite and remain unchanged by renormalization.
Using the results for ${\rm Im}i{\cal I}^{\rm vac}_{\rm sun}(M^2;m_1,m_2,m_3)$
we can easily evaluate \eq{sunsetvac-Re} numerically. 
After ${\rm Re}i\tilde{\cal I}^{\rm vac}_{\rm sun}(m_{phys}^2;m_1,m_2,m_3)_{11}$ is subtracted,
\eq{sunsetvac-Re} coincides with that in modified minimal subtraction scheme up to an irrelevant overall factor.

\subsection{$T$-dependent part}

We rewrite the integrand of the first term in the $T$-dependent part 
\eq{sunset1} so that
it becomes the same form as that in the second term, 
{\it i.e.} (a delta function)$\times$(some functions).
This can be done by recombining the two factors in the integrand 
of $F^{(2)}(k-p;m_2,m_3)+(2\leftrightarrow 3)$ and $F^{(3)}(k-p;m_2,m_3)$,
$T$-independent part and $T$-dependent part of $i\Delta_{11}^F$. 
The result is as follows:
\bey
&&{\cal I}^{\rm th}_{\rm sun}(k;m_1,m_2,m_3)_{11}\cr
&=&\int\mesr{p}\left\{
n(p_0)2\pi\delta(p^2-m_1^2)
\left[I^{(2)}(k-p;m_2,m_3)
+\left(F^{(2)}(k-p;m_2,m_3)+(2\leftrightarrow 3)\right)
\right.\right.\cr
&&\left.+F^{(3)}(k-p;m_2,m_3)\right]\cr
&&+n(p_0)2\pi\delta(p^2-m_2^2)\left[I^{(2)}(k-p;m_1,m_3)+F^{(2)}(k-p;m_1,m_3)\right]\cr
&&\left.
+n(p_0)2\pi\delta(p^2-m_3^2)I^{(2)}(k-p;m_1,m_2)
\right\}.
\label{sunset2}
\eey
When we put $k=(k_0,{\bf 0})$,
\eq{sunset2} is reduced to 
\bey
&&{\cal I}^{\rm th}_{\rm sun}(k_0,{\bf 0};m_1,m_2,m_3)_{11}\cr
&=&\frac{1}{4\pi^2}\sum_{\tau=\pm}\left\{\int_{m_1}^{\infty}{\rm d}\omega
n(\omega)\sqrt{\omega^2-m_1^2}\left[I^{(2)}(k_0+\tau \omega,\sqrt{\omega^2-m_1^2};m_2,m_3)\right.\right.\cr
&&
+\left(F^{(2)}(k_0+\tau \omega,\sqrt{\omega^2-m_1^2};m_2,m_3)+(2\leftrightarrow 3)\right)\cr
&&\left.+F^{(3)}(k_0+\tau \omega,\sqrt{\omega^2-m_1^2};m_2,m_3)\right]\cr
&&+\int_{m_2}^{\infty}{\rm d}\omega n(\omega)\sqrt{\omega^2-m_2^2}
\left[I^{(2)}(k_0+\tau \omega,\sqrt{\omega^2-m_2^2};m_1,m_3)
\right.\cr
&&\left.+F^{(2)}(k_0+\tau \omega,\sqrt{\omega^2-m_2^2};m_1,m_3)\right]\cr
&&\left.
+\int_{m_3}^{\infty}{\rm d}\omega n(\omega)\sqrt{\omega^2-m_3^2}\cdot I^{(2)}(k_0+\tau \omega,\sqrt{\omega^2-m_3^2};m_1,m_2)
\right\},
\label{sunset3}
\eey
Using the expressions of $I^{(2)}$, $F^{(2)}$ and $F^{(3)}$ 
obtained in the previous section, 
we can evaluate the continuous and discontinuous parts of \eq{sunset3} numerically.

\section{Contribution of the thermal sunset diagram to the 
spectral function of the sigma meson at finite temperature}

The purpose of this section is to see how two-loop diagrams
affect observables 
by evaluating the contribution of the thermal sunset diagram 
to $\sigma$ spectral function at finite temperature 
in the ${\cal O}(4)$ linear sigma model.

\begin{figure}[!h]
\begin{center}
\includegraphics{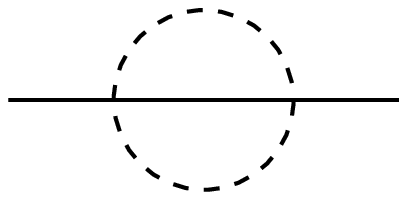}
\caption{The sunset diagram for $\sigma$. 
Solid and dashed lines correspond to $\sigma$ and $\pi$ 
respectively.}
\label{sigmasunset}
\end{center}
\end{figure}

It is known that naive perturbation theory breaks down at $T\neq 0$ 
and that resummation of higher orders is necessary \Refs{weinberg}{linde}.
We adopt here a resummation technique called
optimized perturbation theory (OPT) \Ref{chiku}.
We first briefly review the procedure of OPT applied to 
${\cal O}(4)$ linear sigma model.
The original linear sigma model Lagrangian is as follows:
\ben
{\cal L}=\frac{1}{2}\left[(\partial_{\mu}\phi_i)^2-\mu^2\phi_i^2\right]
-\frac{\lambda}{4!}(\phi_i^2)^2+h\phi_0+{\rm counter\,\,terms},
\een
where $\phi_i=(\sigma,{\vec\pi})$ and $h\phi_0$ being the explicitly
symmetry breaking term.
For the renormalized couplings $\mu^2$, $\lambda$ and $h$ and 
the renormalization point $\kappa$
we use the values determined in \Ref{chiku}:
$\mu^2=-(283\,{\rm MeV})^2$, $\lambda=73.0$, $h=(123\,{\rm MeV})^3$, $\kappa=255\,{\rm MeV}$.

In OPT one adds and subtracts a new mass term with the mass $m$ 
to the Lagrangian.
Thus, we have
\begin{eqnarray} 
{\cal  L}&=&\frac{1}{2}[(\partial_\mu\phi_i)^2-m^2\phi_i^2]
    +\frac{1}{2}\chi\phi_i^2-\frac{\lambda}{4!}(\phi_i^2)^2+h\phi_0
+\mbox{(counter term)},
\eey
where $\chi\equiv m^2-\mu^2$. 
The idea of OPT is reorganization of perturbation theory:
one treats the added one as a tree-level mass term while the subtracted one 
as perturbation.

When the spontaneous symmetry breaking takes place, 
tree level masses of $\pi$ and $\sigma$ read, respectively,
\begin{equation}
m_{0\pi}^2=m^2+\frac{\lambda}{6}\xi^2\mbox{,~~~~~~~~~}
m_{0\sigma}^2=m^2+\frac{\lambda}{2}\xi^2,
\end{equation}
where $\xi$ is the vacuum expectation value of $\sigma$ 
and determined by the stationary condition
for the thermal effective action $V(\xi,T,m^2)$ \Ref{chiku}:
\begin{equation}
    \frac{\partial V(\xi,T,m^2)}{\partial\xi}=0.
\end{equation}
Note that the derivative with respect to $\xi$ does not act on $m^2$.

If Green's functions are calculated in all orders in OPT, 
they should not depend on the arbitrary mass, $m$. 
However, if one truncates perturbation series at a certain order 
they depend on it.
One can determine this arbitrary parameter so that the correction terms
are as small as possible.
We adopt the following condition \Ref{chiku}:
\begin{equation}
{\Pi}_{\pi}(k^2=m_{0\pi}^2)+{\Pi}_{\pi}(k=0;T)=0,
\label{eq:FAC}
\end{equation}
where the first and second terms are
respectively $T$-independent part and $T$-dependent part of 
the one-loop self-energy of $\pi$.

Let us now turn to the discussion on the spectral function of $\sigma$
defined by \eq{rho-RTF}:
\ben
\rho_{\sigma}(k_0,{\bf k})
=-\frac{1}{\pi}\frac{{\rm Im}\bar{\Pi}_{\sigma}(k_0,{\bf k})}
{(k^2-m_{0\sigma}^2-{\rm Re}\bar{\Pi}_{\sigma}(k_0,{\bf k}))^2
+({\rm Im}\bar{\Pi}_{\sigma}(k_0,{\bf k}))^2}.
\een
Here $\bar{\Pi}_{\sigma}(k_0,{\bf k})$ is the self-energy of $\sigma$.
Its real and imaginary parts are related with (1,1) component, 
${\Pi}_{\sigma}^{11}$, {\it via} \eqs{rel-SE1}{rel-SE2}.

As was already mentioned, the spectral function at one-loop order was studied by Chiku and Hatsuda [\ref{chiku}].
We want to see how the spectral function at one-loop order 
is modified by adding the thermal sunset diagram.
Thus, we take
\ben
\Pi_{\sigma}^{11}(k_0,{\bf k})
=\Pi_{\sigma}^{11}(k_0,{\bf k})_{\rm 1-loop}
+\Pi_{\sigma}^{11}(k_0,{\bf k})_{\rm sun}.
\een
The first term is the renormalized one-loop self-energy calculated in [\ref{chiku}].
The second term is the renormalized sunset diagram 
depicted in Fig.\ref{sigmasunset}  
and given by
\bey
\Pi_{\sigma}^{11}(k_0,{\bf k})_{\rm sun}
&=&-\frac{\lambda^2}{6}
i{\cal I}_{\rm sun}(k_0,{\bf k};m_{0\pi},m_{0\pi},m_{0\sigma})_{11}.
\eey

We show $\rho_{\sigma}(k_0,{\bf k}={\bf 0})$ 
at $T=200\,{\rm MeV}$ and $T=145\,{\rm MeV}$ in Fig.\ref{rho}.
\begin{figure}[h!]
\begin{center}
\rotatebox{-90}
{\includegraphics[height=15cm,width=27cm,keepaspectratio]{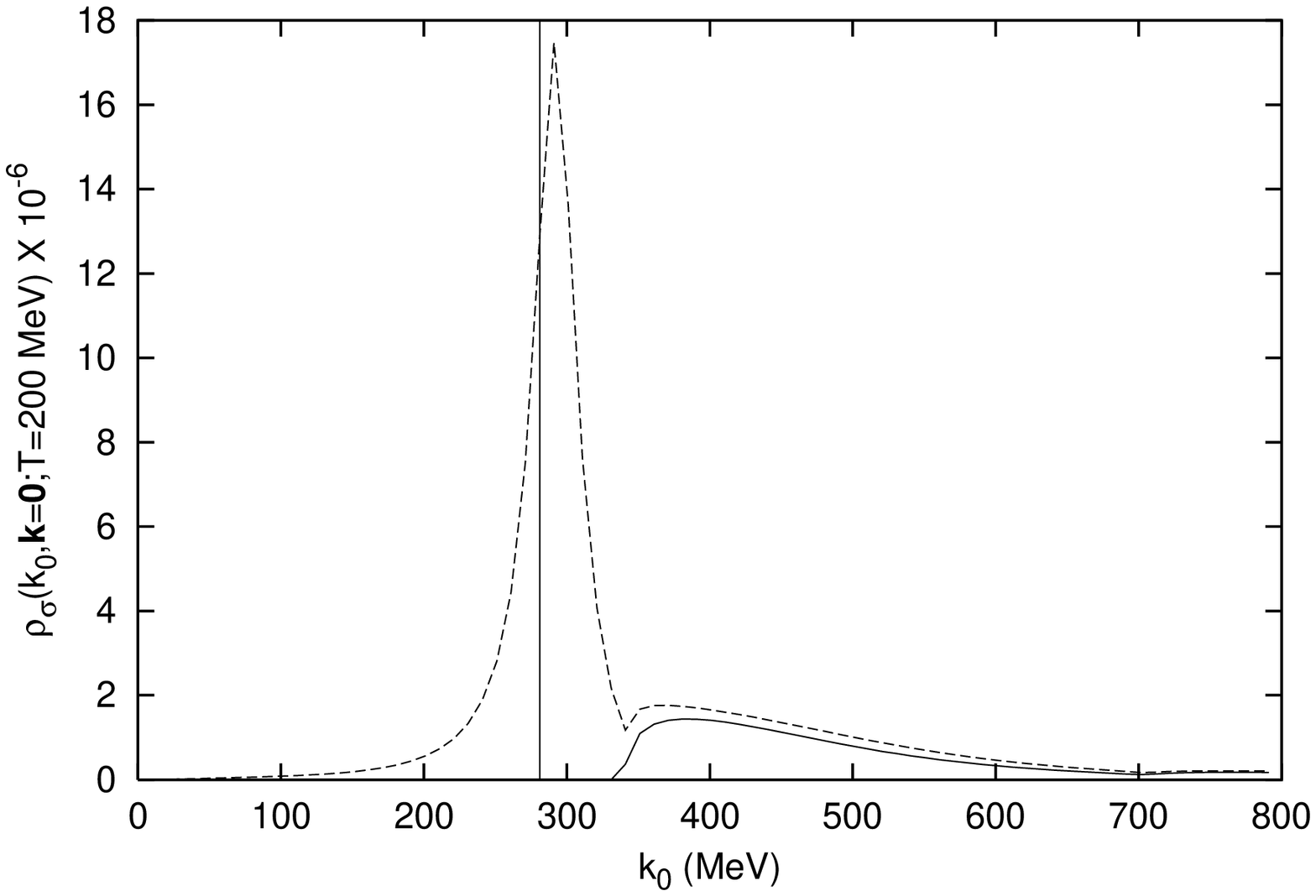}}
\rotatebox{-90}
{\includegraphics[height=15cm,width=27cm,keepaspectratio]{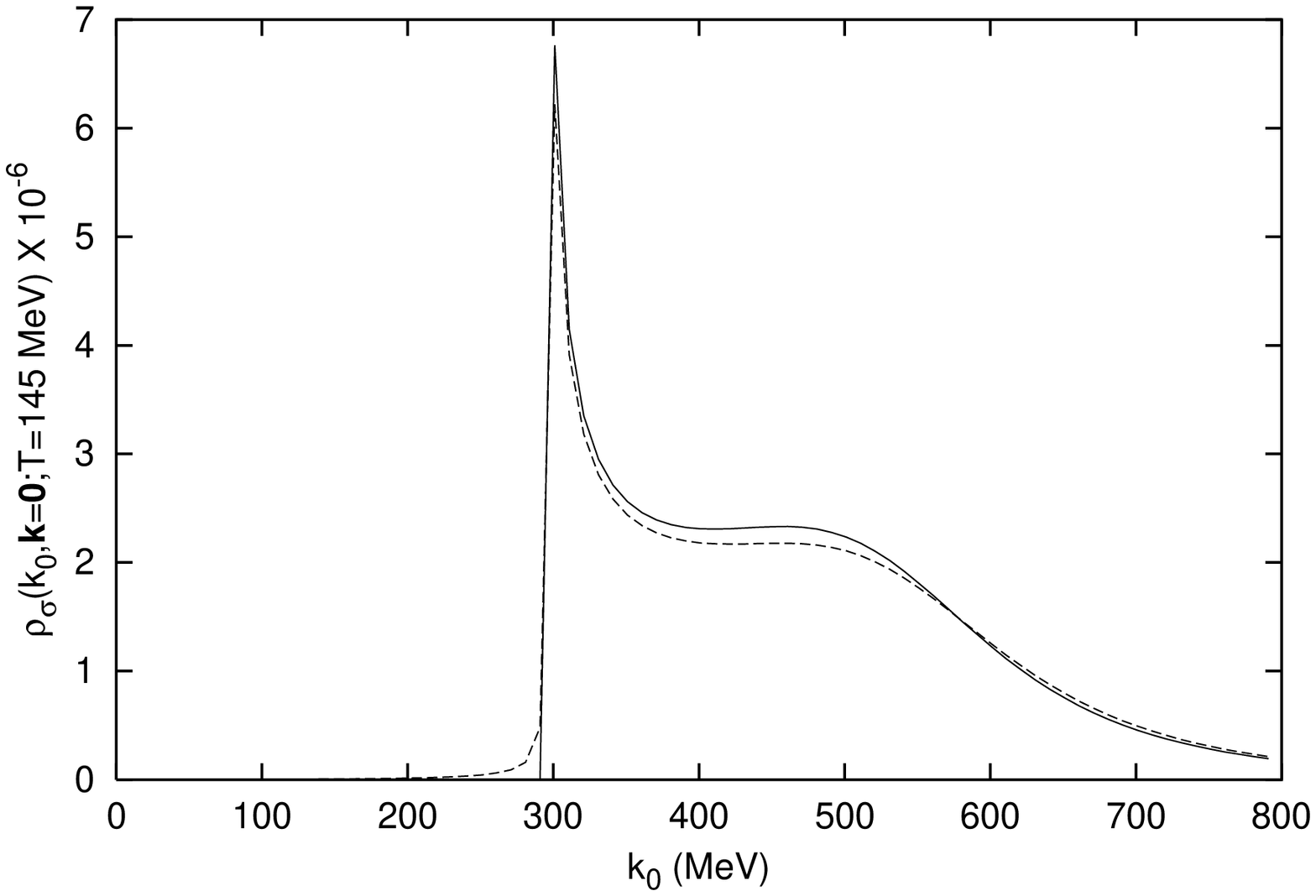}}
\caption{Spectral function of $\sigma$
$\rho_{\sigma}(k_0,{\bf k}={\bf 0})$ at $T=200\,{\rm MeV}$
    (upper panel) and $T=145\,{\rm MeV}$ (lower panel).
Solid line corresponds to $\rho_{\sigma}$ at one-loop order and
dashed line to that with one-loop self-energy 
and the sunset diagram.}
\label{rho}
\end{center}
\end{figure}
At $T=200\,{\rm MeV}$, the spectral function at one-loop order
consists of a $\delta$-function peak for $\sigma$ and a continuum.
By including the sunset diagram $\sigma$ acquires 
a width of the order of $10\,{\rm MeV}$.
At lower temperature, $T=145\,{\rm MeV}$, 
an enhancement of the spectrum near the threshold is observed
at one-loop order. When we include the sunset diagram,
this feature is not lost.

Let us discuss the above results. 
At high temperature ($T=200\,{\rm MeV}$),
the mass of $\sigma$ is smaller 
than twice the pion mass
and the decay $\sigma\rightarrow\pi\pi$ is forbidden.
As a result, within the one-loop calculation $\sigma$ has zero width.
However, at finite temperature $\sigma$ can interact with thermal particles 
in heat bath and change into other states. 
Among such processes, those which are taken into account 
by including the sunset diagram are represented by Fig.\ref{decaysunset}
with $(\Phi,\phi_1,\phi_2,\phi_3)$ assigned to, 
for example, $(\sigma,\sigma,\pi,\pi)$.
The processes which correspond to (a) and (b) in Fig.\ref{decaysunset} 
are possible for $k_0>2m_{0\pi}+m_{0\sigma}$.
This affect the spectrum at high energy.
(c) and (d) in Fig.\ref{decaysunset} drop off for positive $k_0$.
(e) and (f) affect the spectrum at low energy
since they are possible for $k_0<m_{0\sigma}-2m_{0\pi}$.
The processes which correspond to
(g) and (h) in Fig.\ref{decaysunset} are shown in
Fig.\ref{decaysigmasunset}.
They are the most important and give a finite width to $\sigma$
since they are allowed at arbitrary positive $k_0$.
However, we observe that their effects at lower temperature 
($T=145\,{\rm MeV}$) are small.
The reason is traced back to \eq{imsunset2}.
The term representing the probabilities for (g) and (h) 
in Fig.\ref{decaysunset} is the fourth term.
That integral at lower temperature is suppressed 
due to the statistical weight
since at lower temperature the masses of $\sigma$ and $\pi$ are large.

Finally we note that, as a consequence of the non-vanishing
discontinuous part of the sunset diagram,
the spectral function is also non-vanishing in the all range of $k_0$.

\begin{figure}[h]
\begin{center}
\includegraphics{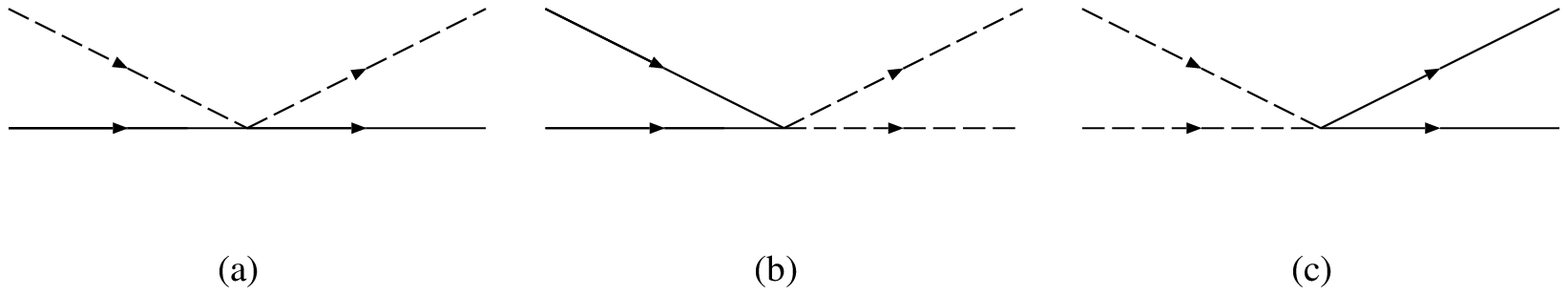}
\caption{The processes which are allowed at all positive $k_0$
contained in the discontinuous part of the sunset diagram 
for $\sigma$ (Fig.\ref{sigmasunset}).}
\label{decaysigmasunset}
\end{center}
\end{figure}

\section{Summary and perspective}
We have studied the sunset diagram for scalar field theories at finite temperature
in the real time formalism.
We have explained how we can reduce it to an expression written in terms of
one-loop self-energy integrals,
which can be easily evaluated numerically.
We have also discussed what physical processes are contained 
in the discontinuous part of the diagram.
We have found that there exist processes 
which occur only at finite temperature and some of them 
are allowed at arbitrary energy.
As a result, the discontinuous part of the sunset diagram
is non-vanishing in all the energy region,
which is a remarkable feature at finite temperature and manifests itself at two-loop order.

As an application of the result,
we have demonstrated how the spectral function of $\sigma$ at finite $T$ 
at one-loop order is modified when we include the thermal sunset diagram.
At high temperature, where $\sigma\rightarrow\pi\pi$ is forbidden,
$\sigma$ acquires a finite width of the order of $10\,{\rm MeV}$
due to collisions with thermal particles in the heat bath
while $\sigma$ does not have a width at one-loop order.
At lower temperature the spectrum is almost unchanged.  

Finally we comment on the effect of other two-loop diagrams on the spectral function. 
We have seen that the threshold enhancement, 
which was first found in the one-loop calculation,
is retained if we include the sunset diagram.
However, in the present calculation
the effect of the thermal width of $\pi$ in the $\sigma\rightarrow\pi\pi$ decay
is not included.
This is taken into account by including the diagrams 
such as shown in Fig.\ref{other2loop},
in which an internal $\pi$ changes into $\sigma$ absorbing a thermal $\pi$ in the heat bath.

In the one-loop calculation, $\pi$ has a width, $\Gamma_\pi=50\,{\rm MeV}$, at the temperature, $T=145\,{\rm MeV}$, at which the threshold enhancement is observed for $\sigma$.
If we include this effect as a constant complex mass shift for $\pi$ in the one-loop self-energy for $\sigma$,
we expect $\sigma$ to acquire a width twice as that for $\pi$, i.e. $\Gamma_\sigma=2\Gamma_\pi$ [\ref{nishi-compmass}].
This implies that when the two-loop diagrams such as Fig.\ref{other2loop} are included the spectral function for $\sigma$ would be significantly modified with the width of about $100\,{\rm MeV}$.
The calculations of those diagrams are now in progress [\ref{nishi-2loop}].

\begin{figure}[b]
\begin{center}
\includegraphics[width=5cm]{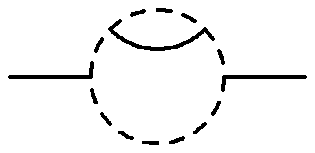}
\caption{two-loop self-energy diagram in which internal $\pi$ changes into $\sigma$ by
absorbing thermal $\pi$ in the heat bath.}
\label{other2loop}
\end{center}
\end{figure}

\acknowledgments
The authors would like to thank M. Ohtani for useful discussion.
This work was partially supported by Grants-in-Aid of the Japanese Ministry
of Education, Science, Sports, Culture and Technology (No.06572).
\clearpage
\appendix
\section{Calculation of ${\rm Im} iF^{(2)}(p;m_1,m_2)$ and $F^{(3)}(p;m_1,m_2)$}
\label{cond-disc}

In this appendix, we derive Eqs.(\ref{ImiF2-1})-(\ref{ImiF2-3}) 
and Eqs.(\ref{F3-1})-(\ref{F3-3}).

From \eq{F2} we obtain
\bey
{\rm Im}iF^{(2)}(p;m_1,m_2)
&=&\frac{1}{2}\int\mesr{k}2\pi\delta((p+k)^2-m_1^2)
n(k_0)2\pi\delta(k^2-m_2^2)
\cr
&=&
\frac{1}{4\pi}
\int_{-\infty}^{\infty}dk_0 \int_{0}^{\infty}d|{\bf k}||{\bf k}|^2
n(k_0)\delta(k^2-m_2^2)\cr
&&\times
\int_{-1}^{1}d{\rm cos}\theta
\delta(p^2+2p_0 k_0-2|{\bf p}||{\bf k}|{\rm cos}\theta+k^2-m_1^2).
\eey
${\rm Im}iF^{(2)}(p;m_1,m_2)$ receives contribution from $k$ which satisfies
\ben
\frac{|p^2+2p_0 k_0+m_2^2-m_1^2|}{2|{\bf p}||{\bf k}|}<1
\label{cond1}
\een
and $k_0^2={\bf k}^2+m_2^2$.
We make a square of \eq{cond1} and obtain
\bey
4p_0 k_0 (p^2+m_2^2-m_1^2)<-(p^2+m_2^2-m_1^2)^2-4p_0^2 m_2^2-4p^2\absk^2.
\label{cond2}
\eey
There are three cases when \eq{cond2} holds:
\begin{enumerate}
\item[A.]
LHS$>0$ and RHS$>0$, (RHS)$^2-$(LHS)$^2>0$.
\item[B.]
LHS$<0$ and RHS$<0$, (RHS)$^2-$(LHS)$^2<0$.
\item[C.]
LHS$<0$ and RHS$>0$.
\end{enumerate}
First, we calculate (RHS)$^2-$(LHS)$^2$.
\bey
&&({\rm RHS})^2-({\rm LHS})^2\cr
&=&16p^4\absk^4+\left\{8p^2\left[(p^2+m_2^2-m_1^2)^2+4p_0^2 m_2^2\right]
-16(p^2+m_2^2-m_1^2)^2 p_0^2\right]\absk^2\cr
&&+\left[(p^2+m_2^2-m_1^2)^2+4p_0^2 m_2^2\right]^2
-16(p^2+m_2^2-m_1^2)^2p_0^2 m_2^2\cr
&=&16p^4\absk^4+8\left[(p^2-2p_0^2)(p^2+m_2^2-m_1^2)^2+4p^2 p_0^2 m_2^2\right]\absk^2\cr
&&+\left[(p^2+m_2^2-m_1^2)^2-4p_0^2 m_2^2\right]^2
\eey
Whether there exists $\absk^2$ which satisfies $({\rm RHS})^2-({\rm LHS})^2<0$ depends on
the sign of the following expression:
\bey
D&=&16\left[(p^2-2p_0^2)(p^2+m_2^2-m_1^2)^2+4p^2 p_0^2 m_2^2\right]^2
-16p^4\left[(p^2+m_2^2-m_1^2)^2-4p_0^2 m_2^2\right]^2\cr
&=&64p_0^2 \absp^2 (p^2+m_2^2-m_1^2)^2\left[p^2-(m_1+m_2)^2\right]
\left[p^2-(m_1-m_2)^2\right].
\eey
Thus, for $(m_1-m_2)^2<p^2<(m_1+m_2)^2$
$$
({\rm RHS})^2-({\rm LHS})^2>0,
$$
for $p^2<(m_1-m_2)^2$ or $(m_1+m_2)^2<p^2$
\begin{eqnarray*}
&&\absk^2<\absk_{-}^2\quad{\rm or}\quad\absk^2>\absk_{+}^2
\quad\Longleftrightarrow\quad ({\rm RHS})^2-({\rm LHS})^2>0,\cr
&&\absk_{-}^2<\absk^2<\absk_{+}^2
\quad\Longleftrightarrow\quad ({\rm RHS})^2-({\rm LHS})^2<0,
\end{eqnarray*}
where
$\absk_{\pm}^2$ are given by
\bey
\absk_{\pm}^2
&=&\frac{1}{4}\left\{
\sqrt{\left(1+\frac{m_2^2-m_1^2}{p^2}\right)^2}\absp
\pm\sqrt{\left[1-\frac{(m_2+m_1)^2}{p^2}\right]
\left[1-\frac{(m_1-m_2)^2}{p^2}\right]}|p_0|\right\}^2.\cr
&&?
\label{kpm}
\eey

Secondly,
for $p^2>m_1^2-m_2^2$
$$
p_0k_0\gtrless 0 \quad\Longleftrightarrow\quad {\rm LHS}\gtrless 0,
$$
for $p^2<m_1^2-m_2^2$
$$
p_0k_0\gtrless 0 \quad\Longleftrightarrow\quad {\rm LHS}\lessgtr 0.
$$

Thirdly, for $p^2>0$
$$
{\rm RHS}<0
$$
for $p^2<0$
$$
\absk^2\gtrless\frac{(p^2+m_2^2-m_1^2)^2+4p_0^2 m_2^2}{-4p^2}\equiv \absk_0^2
\quad\Longleftrightarrow\quad {\rm RHS}\gtrless 0
$$

Therefore, the conditions for the above three cases are respectively given by
\begin{enumerate}
\item[A.]
$p^2<0$, $p_0 k_0(p^2+m_2^2-m_1^2)>0$ and $\absk^2>\absk_{+}^2$.
\item[B.]
$p^2>(m_1+m_2)^2$, $p_0 k_0(p^2+m_2^2-m_1^2)<0$ and $\absk_{-}^2<\absk^2<\absk_{+}^2$,

$0<p^2<(m_1-m_2)^2$ and $p_0 k_0(p^2+m_2^2-m_1^2)<0$ and $\absk_{-}^2<\absk^2<\absk_{+}^2$,

$p^2<0$, $p_0 k_0(p^2+m_2^2-m_1^2)<0$ and $\absk_{-}^2<\absk^2<\absk_0^2$.
\item[C.]
$p^2<0$, $p_0 k_0(p^2+m_2^2-m_1^2)<0$ and $\absk^2>\absk_0^2$.
\end{enumerate}
The case C can be combined with the third of the case B as
\begin{enumerate}
\item[]
$p^2<0$, $p_0 k_0(p^2+m_2^2-m_1^2)<0$ and $\absk^2>\absk_{-}^2$.
\end{enumerate}

\benmrt
\item
For $p^2>(m_1+m_2)^2$ or $0<p^2<(m_1-m_2)^2$,
\bey
{\rm Im}iF^{(2)}(p;m_1,m_2)
&=&\frac{1}{8\pi\absp}\int_{\absk_-}^{\absk_+}d\absk \frac{\absk}{\omega_k}n(\omega_k)\cr
&=&\frac{1}{8\pi\absp}\int_{\omega_-}^{\omega_+}{\rm d}\omega n(\omega)\cr
&=&\frac{1}{16\pi|{\bf p}|}
\frac{1}{\beta}
\ln\left|\frac{1-{\rm e}^{-\beta\omega_+}}
{1-{\rm e}^{-\beta\omega_-}}\right|.
\eey
\item
For $(m_1-m_2)^2<p^2<(m_1+m_2)^2$,
\bey
{\rm Im}iF^{(2)}(p;m_1,m_2)&=&0.
\eey
\item
For $p^2<0$,
\bey
{\rm Im}iF^{(2)}(p;m_1,m_2)
&=&\frac{1}{8\pi\absp}
\left(\int_{\absk_-}^{\infty}+\int_{\absk_+}^{\infty}\right)
d\absk \frac{\absk}{\omega_k}n(\omega_k)\cr
&=&\frac{1}{8\pi\absp}
\left(\int_{\omega_-}^{\infty}+\int_{\omega_+}^{\infty}\right)
{\rm d}\omega n(\omega)\cr
&=&\frac{1}{16\pi|{\bf p}|}\frac{-1}{\beta}
\ln\left|(1-{\rm e}^{-\beta\omega_+})(1-{\rm e}^{-\beta\omega_-})\right|.
\eey
\eenmrt

where
\ben
\omega_{\pm}
=\frac{1}{2}\left|
\sqrt{\left(1+\frac{m_2^2-m_1^2}{p^2}\right)^2}|p_0|
\pm\sqrt{\left[1-\frac{(m_2+m_1)^2}{p^2}\right]
\left[1-\frac{(m_1-m_2)^2}{p^2}\right]}\absp\right|
\een

Similarly,
\benmrt
\item
For $p^2>(m_1+m_2)^2$,
\bey
{\rm Im}iF^{(3)}(p;m_1,m_2)
&=&\frac{1}{8\pi\absp}\int_{\absk_-}^{\absk_+} d\absk\frac{\absk}{\omega_k}
n(\omega_k)n(|p_0|-\omega_k)\cr
&=&\frac{1}{8\pi\absp}\int_{\omega_-}^{\omega_+}{\rm d}\omega 
n(\omega)n(|p_0|-\omega)\cr
&=&\frac{1}{8\pi\absp\beta}\frac{1}{{\rm e}^{\beta|p_0|}-1}
{\rm ln}\left|\frac{1-{\rm e}^{-\beta\omega_+}}{1-{\rm e}^{-\beta\omega_-}}
\frac{{\rm e}^{\beta(|p_0|-\omega_-)}-1}
     {{\rm e}^{\beta(|p_0|-\omega_+)}-1}\right|.
\eey
\item
For $(m_1-m_2)^2<p^2<(m_1+m_2)^2$,
\bey
{\rm Im}iF^{(3)}(p;m_1,m_2)&=&0.
\eey
\item
For $0<p^2<(m_1-m_2)^2$,
\bey
{\rm Im}iF^{(3)}(p;m_1,m_2)
&=&\frac{1}{8\pi\absp}\int_{\absk_-}^{\absk_+} d\absk\frac{\absk}{\omega_k}
n(\omega_k)n(|p_0|+\omega_k)\cr
&=&\frac{1}{8\pi\absp}\int_{\omega_-}^{\omega_+}{\rm d}\omega 
n(\omega)n(|p_0|+\omega)\cr
&=&\frac{1}{8\pi\absp\beta}\cdot\frac{1}{{\rm e}^{\beta|p_0|}-1}\cr
&&\times\left[
{\rm ln}\left|\frac{1-{\rm e}^{-\beta\omega_+}}
                   {1-{\rm e}^{-\beta\omega_-}}\right|
-{\rm e}^{\beta|p_0|}
{\rm ln}\left|\frac{1-{\rm e}^{-\beta(|p_0|+\omega_+)}}
                   {1-{\rm e}^{-\beta(|p_0|+\omega_-)}}\right|
\right].
\eey
\item
For $p^2<0$,
\bey
&&{\rm Im}iF^{(3)}(p;m_1,m_2)\cr
&=&\frac{1}{8\pi\absp}\left(
\int_{\absk_-}^{\infty}d\absk\frac{\absk}{\omega_k}n(\omega_k)n(|p_0|+\omega_k)
+\int_{\absk_+}^{\infty}d\absk\frac{\absk}{\omega_k}n(\omega_k)n(|p_0|-\omega_k)
\right)
\cr
&=&\frac{1}{8\pi\absp}\left(
\int_{\omega_-}^{\infty}{\rm d}\omega n(\omega)n(|p_0|+\omega)
+\int_{\omega_+}^{\infty}{\rm d}\omega n(\omega)n(|p_0|-\omega)
\right)\cr
&=&\frac{1}{8\pi\absp\beta}\left\{\frac{1}{e^{-\beta|p_0|}-1}
\left[
-{\rm ln}|1-{\rm e}^{-\beta\omega_+}|
+{\rm e}^{-\beta|p_0|}{\rm ln}|1-{\rm e}^{-\beta(\omega_{+}-|p_0|)}|
\right]\right.\cr
&&
\left.
+\frac{1}{e^{\beta|p_0|}-1}\left[
-{\rm ln}|1-{\rm e}^{-\beta\omega_-}|
+{\rm e}^{\beta|p_0|}{\rm ln}|1-{\rm e}^{-\beta(\omega_{-}+|p_0|)}|
\right]
\right\}.
\eey
\eenmrt

\section{Physical meaning of $\absk_{\pm}$}
In this appendix we explain the physical meaning of $\absk_{\pm}$ defined by \eq{kpm}.

Let us first consider the case $p^2>(m_1+m_2)^2$ or $0<p^2<(m_1-m_2)^2$.
For simplicity we suppose $p_0>0$.
Since internal particles are on shell, in the center-of-mass frame we have
\ben
p^2=\left(\sqrt{m_1^2+\absk^2}\pm\sqrt{m_2^2+\absk^2}\right)^2,
\een
where $+$ is for $p^2>(m_1+m_2)^2$ and $-$ is for $0<p^2<(m_1-m_2)^2$.
In either case
\ben
\absk^2=\frac{[p^2-(m_1+m_2)^2][p^2-(m_1-m_2)^2]}{4p^2}.
\een
If we boost the system to the positive z-direction, the external and internal momenta become
\bey
&&\left\{\begin{array}{ll}
p_0'=p_0 {\rm cosh}\theta,&{}\\
{\bf p}_{\perp}=0,\quad p_z'=p_0 {\rm sinh}\theta&(\theta>0),
\end{array}
\right.\\
&&\left\{\begin{array}{l}
k_0'=k_0 {\rm cosh}\theta+k_z{\rm sinh}\theta,\\
{\bf k'}_{\perp}={\bf k}_{\perp},\quad k_z'=k_z{\rm cosh}\theta+k_0 {\rm sinh}\theta.
\end{array}
\right.
\eey
Since $-\absk<k_z<\absk$,
\ben
|{\bf k}'|^2_{-}\leq|{\bf k'}|^2\leq|{\bf k}'|^2_{+},
\een
where
\bey
|{\bf k}'|^2_{\pm}&=&(\pm\absk{\rm cosh}\theta+|k_0| {\rm sinh}\theta)^2\cr
&=&\left(\pm\absk\frac{p_0'}{p_0}+|k_0|\frac{p_z'}{p_0}\right)^2\cr
&=&\frac{1}{4}
\left\{\pm\sqrt{\left[1-\frac{(m_1+m_2)^2}{p^2}\right]\left[1-\frac{(m_1-m_2)^2}{p^2}\right]}p_0'+\left|1+\frac{m_2^2-m_1^2}{p^2}\right||{\bf p}'|\right\}^2.
\eey
This is nothing but \eq{kpm}.
Therefore, for $p^2>(m_1+m_2)^2$ or $0<p^2<(m_1-m_2)^2$, $\absk_+$ and $\absk_-$
are respectively the maximum and minimum values of $\absk$
such that the internal particles are on-shell.

Let us next consider the case $p^2<0$.
In the Breit frame, $p=(0,{\bf p})$, with ${\bf p}$ in the positive z-direction we write the momentum of the particle with mass $m_2$ by $k=(k_0,{\bf k})$ where $k_0$ can be positive or negative. 
Then, the momentum of the particle with mass $p+k=(k_0,{\bf p}+{\bf k})$. 
Since internal particles are on shell, we have
\ben
k_0^2=m_2^2+k_z^2+|{\bf k}_{\perp}|^2=m_1^2+(p_z+k_z)^2+|{\bf k}_{\perp}|^2.
\een
From this equation we obtain
\bey
k_z&=&-\frac{p_z^2+m_1^2-m_2^2}{2p_z}
=\frac{-p^2+m_1^2-m_2^2}{2\sqrt{-p^2}},\cr
k_0^2
&=&\frac{[-p^2+(m_1+m_2)^2][-p^2+(m_1-m_2)^2]}{-4p^2}^2+|{\bf k}_{\perp}|^2,
\label{komega}
\eey
When boosting the frame to the z-direction, we suppose the momenta become
\bey
&&p'=(p_0',{\bf p}'),\quad k'=(k_0',{\bf k}'),\quad p'+k'=(p_0'+k_0',{\bf p}'+{\bf k}').
\eey
Then, $p'$ and $k'$ are given by
\bey
&&\left\{
\begin{array}{l}
p_0'=\absp{\rm sinh}\theta,\\
|{\bf p}'|=\absp{\rm cosh}\theta,
\end{array}
\right.
\label{p'}\\
&&\left\{\begin{array}{l}
k_0'=k_0 {\rm cosh}\theta+k_z{\rm sinh}\theta,\\
{\bf k'}_{\perp}={\bf k}_{\perp},\quad k_z'=k_z{\rm cosh}\theta+k_0 {\rm sinh}\theta.\label{k'}
\end{array}
\right.
\eey
Using Eqs.(\ref{komega}), (\ref{p'}) and (\ref{k'}), we obtain
\ben
|{\bf k}'|^2=(k_z{\rm cosh}\theta+k_0 {\rm sinh}\theta)^2+|{\bf k}_{\perp}|^2.
\een
$|{\bf k}'|^2$ has a minimum when $|{\bf k}_{\perp}|^2=0$ but not a maximum.
Depending on the sign of $k_0$, the minimum value is given by
\bey
|{\bf k}'|_{\pm}^2
&=&\left(|k_z|\frac{|\bf p'|}{\sqrt{-p^2}}\pm\sqrt{k_z^2+m_2^2}\frac{p'_0}{\sqrt{-p^2}}\right)^2\cr
&=&\frac{1}{4}\left\{\left|1-\frac{m_1^2-m_2^2}{-p^2}\right||{\bf p'}|\pm\sqrt{\left[1+\frac{(m_1+m_2)^2}{-p^2}\right]\left[1+\frac{(m_1-m_2)^2}{-p^2}\right]}p'_0\right\}^2.
\label{absk'}
%
\eey
\eq{absk'} is nothing but \eq{kpm}.
Therefore, in the case of $p^2<0$, both of $\absk_{+}$ and $\absk_{-}$ 
are the minimum values of
$\absk$ such that the internal particles can be on-shell.

\newpage
\baselineskip 24pt
\begin{center}
{\bf References}
\end{center}
\def\labelenumi{[\theenumi]}
\def\Ref#1{[\ref{#1}]}
\def\Refs#1#2{[\ref{#1},\ref{#2}]}
\def\npb#1#2#3{{Nucl. Phys.\,}{\bf B{#1}},\,#2\,(#3)}
\def\npa#1#2#3{{Nucl. Phys.\,}{\bf A{#1}},\,#2\,(#3)}
\def\np#1#2#3{{Nucl. Phys.\,}{\bf{#1}},\,#2\,(#3)}
\def\plb#1#2#3{{Phys. Lett.\,}{\bf B{#1}},\,#2\,(#3)}
\def\prl#1#2#3{{Phys. Rev. Lett.\,}{\bf{#1}},\,#2\,(#3)}
\def\prd#1#2#3{{Phys. Rev.\,}{\bf D{#1}},\,#2\,(#3)}
\def\prc#1#2#3{{Phys. Rev.\,}{\bf C{#1}},\,#2\,(#3)}
\def\prb#1#2#3{{Phys. Rev.\,}{\bf B{#1}},\,#2\,(#3)}
\def\pr#1#2#3{{Phys. Rev.\,}{\bf{#1}},\,#2\,(#3)}
\def\ap#1#2#3{{Ann. Phys.\,}{\bf{#1}},\,#2\,(#3)}
\def\prep#1#2#3{{Phys. Reports\,}{\bf{#1}},\,#2\,(#3)}
\def\rmp#1#2#3{{Rev. Mod. Phys.\,}{\bf{#1}},\,#2\,(#3)}
\def\cmp#1#2#3{{Comm. Math. Phys.\,}{\bf{#1}},\,#2\,(#3)}
\def\ptp#1#2#3{{Prog. Theor. Phys.\,}{\bf{#1}},\,#2\,(#3)}
\def\ib#1#2#3{{\it ibid.\,}{\bf{#1}},\,#2\,(#3)}
\def\zsc#1#2#3{{Z. Phys. \,}{\bf C{#1}},\,#2\,(#3)}
\def\zsa#1#2#3{{Z. Phys. \,}{\bf A{#1}},\,#2\,(#3)}
\def\intj#1#2#3{{Int. J. Mod. Phys.\,}{\bf A{#1}},\,#2\,(#3)}
\def\sjnp#1#2#3{{Sov. J. Nucl. Phys.\,}{\bf #1},\,#2\,(#3)}
\def\pan#1#2#3{{Phys. Atom. Nucl.\,}{\bf #1},\,#2\,(#3)}
\def\app#1#2#3{{Acta. Phys. Pol.\,}{\bf #1},\,#2\,(#3)}
\def\jmp#1#2#3{{J. Math. Phys.\,}{\bf {#1}},\,#2\,(#3)}
\def\cp#1#2#3{{Coll. Phen.\,}{\bf {#1}},\,#2\,(#3)}
\def\epjc#1#2#3{{Eur. Phys. J.\,}{\bf C{#1}},\,#2\,(#3)}
\def\mpla#1#2#3{{Mod. Phys. Lett.\,}{\bf A{#1}},\,#2\,(#3)}
\def\etal{{\it et al.}}
\begin{enumerate}

\divide\baselineskip by 4
\multiply\baselineskip by 3

\item \label{chiku}
S. Chiku and T. Hatsuda, \prd{57}{R6}{1998}; \prd{58}{76001}{1998}
\item \label{schwin}
J. Schwinger, \jmp{2}{407}{1961}.
\item \label{kel}
L.V. Keldysh, JETP(Sov. Phys.)\,{\bf 20},\,1018\,(1964).
\item \label{craig}
R.A. Craig, \jmp{9}{605}{1968}.
\item 
G.-Z. Zhou, Z.-B. Su, B.-L Hao and Lu Yu, \prb{22}{3385}{1980}.
\item
Y. Takahashi and H. Umezawa, \cp{2}{55}{1975}.
\item
G.W. Semenoff and Y. Takahashi, \npb{220[FS8]}{196}{1983}.
\item
l.P. Kadanoff and G. Baym, 
Quantum Statistical Mechanics (Benjamin, New York 1969);
D.C. Langreth, in 1975 NATO ADI on Linear and Non-Linear Electron Transport
in Solids, J.T. Devreese and E. van Boem eds. (Plenum, New York 1976)
\item
R. Mills, Propagators for many-particle systems 
(Gordon and Breach, New york 1969)
\item
 E.M. Lifshitz and L.P. Pitaevskii, Course of theoretical physics, vol.10:
Physical kinetics (Pergamon, New York 1981).
\item
H. Umezawa, H. Matsumoto and M. Tachiki, 
Thermofield Dynamics and Condensed States (North-Holland, Amsterdam, 1982).
\item
A.J. Niemi and G.W. Semenoff, \ap{152}{105}{1984}.
\item \label{NiemiSeme}
A.J. Niemi and G.W. Semenoff, \npb{230}{181}{1984}.
\item \label{LeBellac}
M. Lebellac, {\it Thermal Field Theory}, Cambridge University Press 1996.
\item \label{sem-ume}
G.W. Semenoff and H. Umezawa, \npb{220}{196}{1983}
\item \label{ramond}
See, for example,
P. Ramond, {\it Field Theory, A Modern Primer}  (Benjamin/Cummings, New York, 1981).
\item \label{fujimoto}
Y. Fujimoto, M. Morikawa and M. Sasaki, \prd{33}{590}{1986}.
\item \label{weldon}
H.A. Weldon, \prd{28}{2007}{1983}.
\item \label{post}
P. Post and J.B. Tausk, \mpla{11}{2115}{1996}.
\item \label{gass}
J. Gasser and M.E. Sainio, \epjc{6}{297}{1999}.
\item \label{braaten}
J.O. Andersen and E. Braaten, \prd{62}{45004}{2000}; \prd{51}{6990}{1995}.
\item \label{groote}
S. Groote, J.G. Koelner and A.A. Pivovarov, \plb{443}{269}{1998};
\npb{542}{515}{1999}; \epjc{11}{279}{1999}.
\item \label{weinberg}
S. Weinberg, \prd{9}{3357}{1974};\\
L. Dolan and R. Jackiw, \prd{9}{3320}{1974}.
\item \label{linde}
D.A. Kirzhnits and A.D. Linde, Ann. Phys.\,(N.Y.){\bf{101}},\,195\,(1976).
\item \label{nishi-compmass}
T. Nishikawa, O. Morimatsu, Y. Hidaka and M. Ohtani, in preparation.
\item \label{nishi-2loop}
T. Nishikawa, O. Morimatsu and Y. Hidaka, in preparation.

\end{enumerate}

\end{document}